\documentclass[conference]{IEEEtran}
\IEEEoverridecommandlockouts
\usepackage{dsfont}
\usepackage{layout}
\usepackage{graphicx}
\usepackage{amssymb}
\usepackage{amsmath}
\usepackage{cite}
\usepackage{mathrsfs}
\usepackage[displaymath,mathlines]{lineno}
\usepackage{xcolor}
\usepackage{multirow}
\usepackage{subfig}
\usepackage{algpseudocode}
\usepackage{algorithm,algpseudocode}
\usepackage{algorithmicx}
\usepackage{pbox}
\usepackage{multicol}
\usepackage{lipsum}
\usepackage{amsthm}
\usepackage{lipsum}
\usepackage{epstopdf}
\usepackage{makecell}
\usepackage{diagbox}
\usepackage{enumitem}

\newcommand{\doublewidetilde}[1]{{%
		\mathpalette\double@widetilde{#1}%
	}}
	
\makeatletter

\def\BState{\State\hskip-\ALG@thistlm}
\makeatother

\begin{document}
%
\title{On the Performance of Image Recovery in Massive MIMO Communications}

\author{\IEEEauthorblockN{Phan Thi Kim Chinh\IEEEauthorrefmark{2}, Trinh Van Chien\IEEEauthorrefmark{3}, Tran Manh Hoang\IEEEauthorrefmark{2}, Nguyen Tien Hoa \IEEEauthorrefmark{2}*, and Van Duc Nguyen\IEEEauthorrefmark{2}}
	\IEEEauthorblockA{\IEEEauthorrefmark{2} School of Electronics and Telecommunications, HUST, Hanoi, Vietnam}
	\IEEEauthorblockA{\IEEEauthorrefmark{3} Department of Electrical
		Engineering (ISY), Link\"{o}ping University, SE-581 83 Link\"{o}ping, Sweden}
	\thanks{This research is funded by Vietnam National Foundation for Science and
Technology Development (NAFOSTED) under grant number 102.01-2019.07.
Corresponding author*: hoa.nguyentien@hust.edu.vn }
\thanks{This paper was presented at the $8$-th International Conference on Communications and Electronics (ICCE 2020) .  \copyright 2020 IEEE. Personal use of this material is permitted. Permission from IEEE must be  obtained  for  all  other  uses,  in  any  current  or  future  media,  including reprinting/republishing this material for advertising or promotional purposes, creating new collective  works, for resale or redistribution to  servers or lists, or reuse of any copyrighted component of this work in other works.}
}

\maketitle

\begin{abstract}
Massive MIMO (Multiple Input Multiple Output) has demonstrated as a potential candidate for 5G-and-beyond wireless networks. Instead of using Gaussian signals as most of the previous works, this paper makes a novel contribution by investigating the transmission quality of image data by utilizing the Massive MIMO technology. We first construct a framework to decode the image signal from the noisy received data in the uplink Massive MIMO transmission by utilizing the alternating direction method of multipliers (ADMM) approach. Then, a low-pass filter is exploited to enhance the efficiency of the remaining noise and artifacts reduction in the recovered image. Numerical results demonstrate the necessity of a post-filtering process in enhancing the quality of image recovery.
\end{abstract}


\IEEEpeerreviewmaketitle

\section{Introduction}
Massive MIMO (Multiple Input Multiple Output) has demonstrated as a potential candidate for 5G-and-beyond wireless communications base on its advantages by the intensification of either the spectrum efficiency or energy efficiency \cite{Bjornson2016c, al2018successive}. This disruptive technology is able to make the best of its antenna array to concentrate each beamforming vector to diverse individual users. That permits it to obtain spectral efficiency much better than conventional MIMO systems where each base station is only equipped with a few antennas \cite{Marzetta2010a}. In Massive MIMO communications, each antenna array can be employed in integration to rise transmitted signals gain, which makes for a highly energy-efficient system. The network throughput is ameliorated to satisfy the going up demands of high data rate in situation there are many different type of intelligent elements are associated. However, most of the previous works in Massive MIMO literature only consider non-structure data, e.g., Gaussian signals, to investigate the system performance.

Image and video transmission has been vital problems and had a large number of researches for the intention of enhancing performance efficiency due to higher requirements and increasing mobile data nowadays than before. These days technology is improving and developing which supports communication through the internet for example chatting, calling from a device to others, talking face to face without any hesitation about the distance between users \cite{zhu2018comparative}. For those media applications demand large bandwidth, satisfied delivery technique and efficient routing protocols \cite{antonakoglou2018toward}. Image and video signals occupy large space in memory devices and take long time for transmitting in a wireless channel, thus it requires efficient transmission and reconstruction protocols. Due to the fact that mobile data rates keep rising and more consumers rely upon wireless transmission, the figure for videos and images transmitted will probably continue increasing in 5G-and-beyond systems \cite{zhou20195g}. As the best of our knowledge, there is no investigation on the reconstruction performance of transmitting image data information by utilizing the Massive MIMO technology.

In this paper, we first introduce an uplink Massive MIMO system model where a user is sending image data to the base station instead of Gaussian signals as many previous works. The image data is modulated by utilizing a modulation scheme before broadcasting to radio channels. From the base station side, we formulate and solve an optimization problem, which reconstructs the transmitted image from received noisy measurements. In particular, the solution to this optimization problem is obtained by a low computational complexity algorithm base on the ADMM approach. We observe that there are still noise and artifacts in each recovered image, thus a low-pass filter is used to enhance the reconstruction quality. We then demonstrate the effectiveness of our proposed algorithm by testing different natural images.

\textit{Notations}: We use the lower and upper bold letters for vectors and matrices, respectively.  $\mathcal{CN}(\cdot, \cdot)$ denotes the circularly symmetric complex Gaussian distribution. The Euclidean and $\ell_0$ norm are denoted by $\| \cdot \|$ and $\| \cdot \|_0$, respectively.

\section{Uplink Image Data Transmission}
This section presents an uplink Massive MIMO transmission
model where a user communicates with the base station. We
then formulate an image reconstruction optimization problem.
\subsection{System Model}
We consider the uplink transmission of a Massive MIMO system in which a base station is equipped with $M$ antennas and a $K$-antennas user.\footnote{The framework builded up for a single user Massive MIMO system in this paper can be easily generalized to multi-cell and multi-user scenarios.} We assume that this user is sending an image $\mathbf{X}$, which is divided into $Z$ column vectors
\begin{equation}
\mathbf{X} = \{ \mathbf{x}_1, \ldots, \mathbf{x}_Z\},
\end{equation}
with $\mathbf{x}_z \in \mathbb{C}^K, z = 1, \ldots , Z$. We further define a finite constellation set $\mathcal{O}$, then each vector contains image information that is mapped into a constellation point. Mathematically, the mapping process is formulated as
\begin{equation}
\{ \mathbf{x}_1, \ldots, \mathbf{x}_Z\} \xrightarrow{\mathcal{M}} \{ \mathbf{s}_1, \ldots, \mathbf{s}_Z\},
\end{equation}
where $\mathbf{s}_z$ belongs to the set $\mathcal{O}^K$. It means that this vector includes all of the constellation points obtained from the image information vector $\mathbf{x}_z$. We stress that if we have a given set of received constellation points, the image data can be obtained by using an inverse mapping $\mathcal{M}^{-1}$ as
\begin{equation}
 \{ \mathbf{s}_1, \ldots, \mathbf{s}_Z\}  \xrightarrow{\mathcal{M}^{-1}} \{ \mathbf{x}_1, \ldots, \mathbf{x}_Z\},
\end{equation}
Let us denote $\mathbf{H} \in \mathbb{C}^{M \times K}$ the propagation channel between this user and the base station. The received baseband signal at the base station, $\mathbf{y}_z \in \mathbb{C}^M$, is formulated as
\begin{equation} \label{eq:ReceivedSig}
\mathbf{y}_z =  \sqrt{\rho}\mathbf{H}\mathbf{s}_z + \mathbf{n}_z,
\end{equation}
where $\mathbf{n}_z \sim \mathcal{CN} (\mathbf{0}, \sigma^2 \mathbf{I}_M)$ denotes the circularly symmetric complex Gaussian noise and $\rho$ is the transmit power that user assigns to each transmitted constellation point. We therefore define the ratio $\rho / \sigma^2$ as the signal-to-noise (SNR) level. From the received signal in \eqref{eq:ReceivedSig}, the base station aims at reconstructing the image data information, which has been contaminated by Gaussian noise.

\subsection{Problem Formulation}
 We now formulate an image recovery problem that can minimize the transmission error over wireless environment with the Massive MIMO technology. More precisely, the optimization problem is formulated as
\begin{equation} \label{Problem1}
\begin{aligned}
& \underset{ \mathbf{s}_z }{\mathrm{minimize}} && \frac{1}{2} \| \mathbf{y}_z - \sqrt{\rho} \mathbf{H} \mathbf{s}_z \|_2^2 \\
& \mbox{subject to} && \mathbf{s}_z \in \mathcal{O}^K,
\end{aligned}
\end{equation}
which is a combinatorial problem therefore requiring high computational complexity due to the discrete set $\mathcal{O}$, for example quadrature amplitude modulation (QAM) or phase-shift keying (PSK) modulation \cite{shahabuddin2017admm}.\footnote{Even though problem~\eqref{Problem1} is formulated based on the perfect channel
state information (CSI), the same approach can be extended to handle the
assumption that only imperfect CSI is available if channel estimation errors
are treated as Gaussian noise.} For further manipulation,  we introduce a relaxation by setting up a convex polytope $\mathcal{P}_{\mathcal{O}}$ surrounding the set $\mathcal{O}$, then problem~\eqref{Problem1} is recast as
\begin{equation} \label{Problem2}
\begin{aligned}
& \underset{ \mathbf{s}_z }{\mathrm{minimize}} && \frac{1}{2} \| \mathbf{y}_z - \sqrt{\rho} \mathbf{H} \mathbf{s}_z \|_2^2 \\
& \mbox{subject to} && \mathbf{s}_z \in \mathcal{P}_{\mathcal{O}}^K.
\end{aligned}
\end{equation}
It notices that we can obtain the optimal solution to problem~\eqref{Problem2} since the objective function is quadratic and the feasible set are convex. We present in detail the solution to problem~\eqref{Problem2} in the next section.
\section{Image Recovery}
In this section, we present an application of using ADMM approach to recover the transmitted image from the noisy received data. Due to the correlation among pixels of the image, we then motivate to use a low-pass filter for a refinement process to obtain a better recovered image. 
\subsection{Solution to the Image Recovery Problem}
For further processing, we first recast \eqref{Problem2} into the corresponding unconstrained optimization problem as
\begin{equation} \label{Problem3}
\begin{aligned}
& \underset{ \mathbf{s}_z }{\mathrm{minimize}} && \frac{1}{2} \| \mathbf{y}_z - \sqrt{\rho} \mathbf{H} \mathbf{s}_z \|_2^2 + \mathds{1}(\mathbf{s}_z),
\end{aligned}
\end{equation}
where the indicator function $ \mathds{1}(\mathbf{s}_z)$ is defined by
\begin{equation}
 \mathds{1}(\mathbf{s}_z) = \begin{cases}
 0, & \mbox{if } \mathbf{s}_z \in  \mathcal{P}_{\mathcal{O}}^K,\\
 \infty, & \mbox{if } \mathbf{s}_z \notin  \mathcal{P}_{\mathcal{O}}^K.
 \end{cases}
\end{equation}
In the objective function of \eqref{Problem3}, the first term aligns with a quadratic form while the challenge comes from the second term. Subsequently, by utilizing the alternating direction method of multipliers, we introduce a new variable $\mathbf{u}_z $ and rewrite problem~\eqref{Problem3} to as
\begin{equation} \label{Problem4}
\begin{aligned}
& \underset{ \mathbf{s}_z, \mathbf{u}_z }{\mathrm{minimize}} && \frac{1}{2} \| \mathbf{y}_z - \sqrt{\rho} \mathbf{H} \mathbf{s}_z \|_2^2  + \mathds{1}(\mathbf{u}_z)\\
& \mbox{subject to} && \mathbf{u}_z = \mathbf{s}_z.
\end{aligned}
\end{equation}
Here, we have decomposed the objective function into two terms with the separate optimization variables, which later shows that it has benefits when utilizing ADMM approach to obtain a solution. The augmented Lagrangian is defined for problem~\eqref{Problem4} as
\begin{equation}
\mathcal{L} (\mathbf{s}_z, \mathbf{u}_z, \pmb{\mu}_z) = \frac{1}{2} \| \mathbf{y}_z - \sqrt{\rho} \mathbf{H} \mathbf{s}_z \|_2^2  + \mathds{1}(\mathbf{u}_z) + \frac{\beta_z}{2}\| \mathbf{s}_z- \mathbf{u}_z - \pmb{\mu}_z\|_2^2,
\end{equation}
where $\pmb{\mu}_z$ is the Lagrange multiplier and the positive  regularization parameter $\beta_z$ is fixed in our considered scenario to avoid the problem of numerical instabilities. From the initial values $\{ \mathbf{s}_z^{(0)}, \mathbf{u}_z^{(0)}, \pmb{\mu}_z^{(0)} \}$ and a given parameter $\beta_z$, we hereafter present an ADMM approach to obtain solution problem~\eqref{Problem4} in an iterative manner.
\subsubsection{Update the $\mathbf{s}_z$ sub-problem} From the given optimized results $\mathbf{u}_z^{(n-1)}, \pmb{\mu}_z^{(n-1)}$, iteration~$n$ will update the solution to variable $\mathbf{s}_z$ by solving the sub-problem
\begin{equation} \label{s-subproblem}
 \mathbf{s}_z^{(n)} = \underset{\mathbf{s}_z}{\textrm{argmin}} \, \frac{1}{2} \| \mathbf{y}_z - \sqrt{\rho}\mathbf{H} \mathbf{s}_z \|_2^2 + \frac{\beta_z}{2}\|\mathbf{u}_z^{(n-1)} - \mathbf{s}_z - \pmb{\mu}_z^{(n-1)}\|_2^2.
\end{equation}
We stress that \eqref{s-subproblem} is a standard quadratic problem, its optimal solution can be obtained by computing the first derivative with respect to $\mathbf{s}_z$ and equaling it to zero as
\begin{equation}
\sqrt{\rho}\mathbf{H}^H \left(\mathbf{y}_z - \sqrt{\rho}\mathbf{H} \mathbf{s}_z^{(n)}  \right) + \beta_z \left( \mathbf{u}_z^{(n-1)} - \mathbf{s}_z^{(n)} - \pmb{\mu}_z^{(n-1)} \right) = 0,
\end{equation}
for which the solution to $\mathbf{s}_z$ is found at iteration $n$ as
\begin{equation} \label{eq:szn}
\begin{split}
\mathbf{s}_z^{(n)} =& \left( \rho \mathbf{H}^H \mathbf{H} + \beta_z \mathbf{I}_K \right)^{-1} \times\\
&\left( \sqrt{\rho} \mathbf{H}^H  \mathbf{y}_z + \beta_z \mathbf{u}_z^{(n-1)}- \beta_z \pmb{\mu}_z^{(n-1)} \right).
\end{split}
\end{equation}
The main merits of \eqref{eq:szn} are not only giving the closed-form expression to the solution of $\mathbf{s}_z$ but also revealing the effects of different factors to the current solution such as the propagation channel, the received signal, and the solution from the previous iteration.
\subsubsection{Update the $\mathbf{u}_z$ subproblem} From the given optimized results $\mathbf{s}_z^{(n)}, \pmb{\mu}_z^{(n-1)}$, the solution to $\mathbf{u}_z$ is obtained by solving the sub-problem 
\begin{equation} \label{eq:uzn}
\mathbf{u}_z^{(n)} = \underset{\mathbf{u}_z \in \mathcal{P}_{\mathcal{O}}^K}{\textrm{argmin}} \,  \mathds{1}(\mathbf{u}_z) + \frac{\beta_z}{2}\| \mathbf{u}_z - \mathbf{s}_z^{(n)} - \pmb{\mu}_z^{(n-1)}\|_2^2.
\end{equation}
We obtain the solution $\mathbf{u}_z^{(n)}$ by projecting the current available information $\mathbf{s}_z^{(n)} + \pmb{\mu}_z^{(n-1)}$ onto the convex polytope $\mathcal{P}_{\mathcal{O}}^K$ as
\begin{equation} \label{eq:uznk}
\left[ \mathbf{u}_z^{(n)} \right]_k = \underset{ p \in \mathcal{P}_{\mathcal{O}} }{\textrm{argmin}} \left| \left[\mathbf{s}_z^{(n)} + \pmb{\mu}_z^{(n-1)} \right]_k - p \right|,
\end{equation}
which projects each element of vector $[\mathbf{s}_z^{(n)} + \pmb{\mu}_z^{(n-1)}]_k, \forall k = 1, \ldots, K,$ onto the convex polytope $\mathcal{P}_{\mathcal{O}}$.
\subsubsection{Update the Lagrange multiplier} After obtaining the solution to $\mathbf{s}_z^{(n)}$ and $\mathbf{u}_z^{(n)}$, the solution to $\pmb{\mu}_z$ is updated in the standard way as
\begin{equation} \label{eq:multiplier}
\pmb{\mu}_z^{(n)} \leftarrow \pmb{\mu}_z^{(n-1)} - \alpha \left(\mathbf{u}_z^{(n)} - \mathbf{s}_z^{(n)} \right),
\end{equation}
where $\alpha > 0$ is a suitable selected step-size to ensure the convergence of the Lagrange multiplier along iterations \cite{van2018distributed}. 

The suggested iterative update is implemented until, the variation of the Lagrangian function between the two consecutive iterations does not exceed a threshold:
\begin{equation} \label{eq:Stopping}
\left| \mathcal{L} \left(\mathbf{s}_z^{(n-1)}, \mathbf{u}_z^{(n-1)}, \pmb{\mu}_z^{(n-1)} \right) - \mathcal{L} \left(\mathbf{s}_z^{(n)}, \mathbf{u}_z^{(n)}, \pmb{\mu}_z^{(n)} \right) \right| \leq \epsilon,
\end{equation}
where $\epsilon$ is a given small constant value. The suggested image recovery method is summarized in Algorithm~\ref{Algorithm1}. By using the same methodology as \cite{Shen2012a}, we can point out that the convexity of problem~\eqref{Problem2}  guarantees the convergence of  Algorithm~\ref{Algorithm1}.
\begin{algorithm}[t]
	\caption{Solution to \eqref{Problem4} based on  ADMM.} \label{Algorithm1}
	\textbf{Input}:  Channel matrix $\mathbf{H}$ and received signal $\mathbf{y}_z, \forall z$.
	\begin{itemize}
		\item[1.]  Initial values $\mathbf{s}_z$, $\mathbf{u}_z$, and $\pmb{\mu}_z, \forall z$. Set $n=0$
		\item[2.] \textbf{While} \textit{Stopping criterion \eqref{eq:Stopping} is not satisfied} \textbf{do}
	\begin{itemize}
		\item[2.1] Set $n = n+1$.
		\item[2.2] Update the sub-problem $\mathbf{s}_{z}^{(n)}$ by utilizing \eqref{eq:szn} for given $\mathbf{u}_z^{(n-1)}$ and  $\pmb{\mu}_z^{(n-1)}$.
		\item[2.3] Update the sub-problem $\mathbf{u}_{z}^{(n)}$ by utilizing \eqref{eq:uznk} for given $\mathbf{s}_z^{(n)}$ and  $\pmb{\mu}_z^{(n-1)}$.
		\item[2.4] Update the multiplier $\pmb{\mu}_{z}^{(n)}$ by utilizing \eqref{eq:multiplier} for given $\mathbf{s}_z^{(n)}$, $\mathbf{u}_z^{(n)}$, and  $\pmb{\mu}_z^{(n-1)}$. 
	\end{itemize}
	\textbf{End While}
	\item[3.] Store the constellation points $\mathbf{s}_z^{\ast} = \mathbf{s}_z^{(n)}$.  
	\item[4.] Do the inverse mapping to get the refined image information $ \mathbf{s}_z  \xrightarrow{\mathcal{M}^{-1}} {\mathbf{x}}_z^{\ast}$.
	\item[5.] Repeat Steps $1$, $2$, $3$ and $4$ for all other received signals $\{\mathbf{y}_z \}$ to obtain $\{ \mathbf{x}_z^{\ast} \}$. 
	\item[6.] Restore the refined image $\widetilde{\mathbf{X}} = \left\{ \mathbf{x}_z^{\ast} \right\}$.
	\end{itemize}
	\textbf{Output}: The refined image $\widetilde{\mathbf{X}}$. 
\end{algorithm}
\subsection{Post Image Filtering}
We stress that the recovery problem presented in the previous subsection has not considered the data structure as a prior information, which is of paramount importance in image and  video processing \cite{van2017block}. Consequently, noise and artifacts may still remain, especially at a low SNR level. We now restore all the decoded image vectors into an image such that
\begin{equation}
\widehat{\mathbf{X}}= \{ \hat{\mathbf{x}}_1, \ldots, \hat{\mathbf{x}}_{Z} \},
\end{equation}
then exploit a low pass filter to mitigate noise and artifacts which remained in the decoded image. 
\subsubsection{Gaussian Filter} This filter is widely used in image processing due to its low computational complexity and flexibility in varying the smoothness level \cite{van2013edge}. Mathematically, the filtered image is obtained as
\begin{equation}
\widetilde{\mathbf{X}} = \widehat{\mathbf{X}} \circledast \mathbf{I},
\end{equation}
where $\circledast$ is the convolutional operator and  $\mathbf{I} \in \mathbb{R}^{u \times u}$ is a smoothing kernal with $u$ is the kernel's size. There are several different choices of the smoothing kernels, but this paper focus on the contributions of the Gaussian filter. In particular, the filtering process is implemented by using a two-dimensional Gaussian kernel  which is defined as
\begin{equation}
\mathbf{I}(x,y) = \frac{1}{2\pi \sigma^2} e^{- \frac{x^2 + y^2}{\sigma^2}},
\end{equation}
where each pair $(x,y)$ stands for a coordinate in the kernel with $-u \leq x,y \leq u$ and $\sigma$ is the standard derivation of the distribution. We stress that the smoothness level of a Gaussian filter is controlled by changing the variance $\sigma^2$.
\subsubsection{Path-based Filter} This filtering process is based on the fact that an image contains both local and non-local correlation, while the latter has significant contributions to preserve texture in noisy images. Besides, the patch-based
sparse representation can effectively handles the complex variations which contained in natural images. Mathematically, each path $\ddot{\mathbf{x}}_z$ of the noisy image $\widehat{\mathbf{X}}$ has a sparse vector $\pmb{\alpha}_z$ which is defined over a dictionary $\mathcal{D}$ such that
\begin{equation}
\ddot{\mathbf{x}}_z \approx \mathcal{D} \pmb{\alpha}_z,
\end{equation}
and then the entire image $\widehat{\mathbf{X}}$ can be represented by $\{ \pmb{\alpha}_z\}$. In order to find each sparse vector $\pmb{\alpha}_z$, we formulate and solve the following optimization problem
\begin{equation} \label{Probv2}
\underset{\pmb{\alpha}_z}{\mathrm{minimize}} \, \, \frac{1}{2} \left\| \mathbf{x}_z -\mathcal{D} \pmb{\alpha}_z \right\|_2^2 + \nu_z \| \pmb{\alpha}_z \|_0,
\end{equation}

\begin{figure*}
	\centerline{%
		\setlength\tabcolsep{-2pt} 
		\begin{tabular}{c  c c c} 	
			\subfloat[Lenna, ground truth]{\label{fig1a} \includegraphics[trim=2.6cm 2cm 2.6cm 2cm, clip=true, width=1.8in]{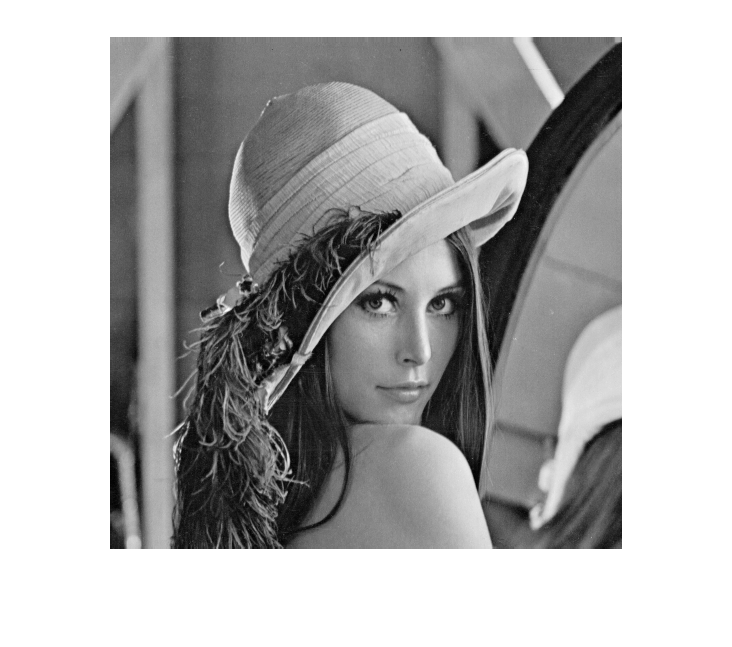}}&
			\subfloat[Barbara, ground truth]{\label{fig1b} \includegraphics[trim=2.6cm 2cm 2.6cm 2cm, clip=true, width=1.8in]{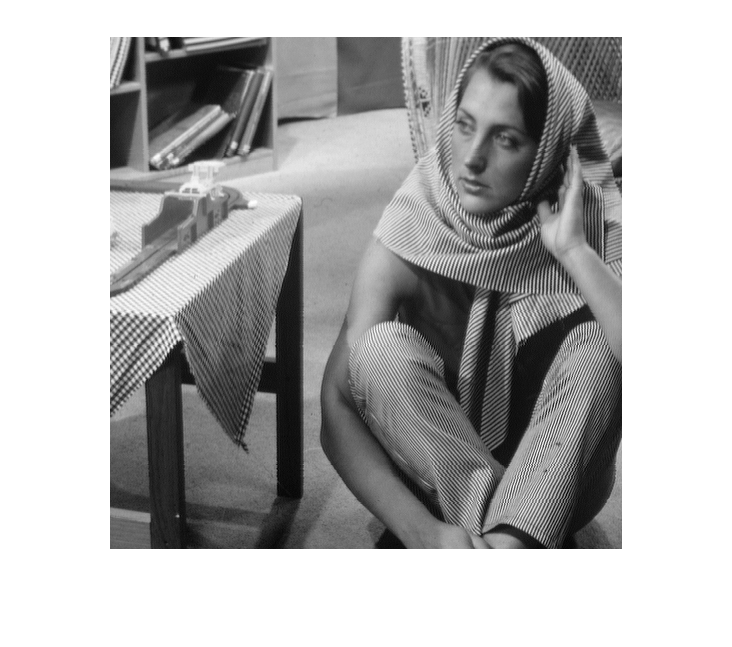}} &
			\subfloat[Mandrill, ground truth]{\label{fig1c} \includegraphics[trim=2.6cm 2cm 2.6cm 2cm, clip=true, width=1.8in]{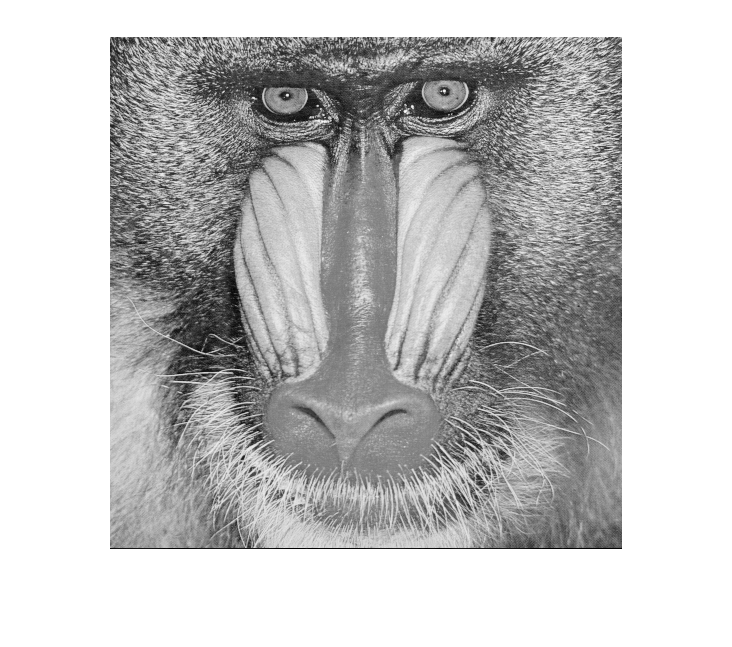}} &
			\subfloat[Peppers, ground truth]{\label{fig1d} \includegraphics[trim=2.6cm 2cm 2.6cm 2cm, clip=true, width=1.8in]{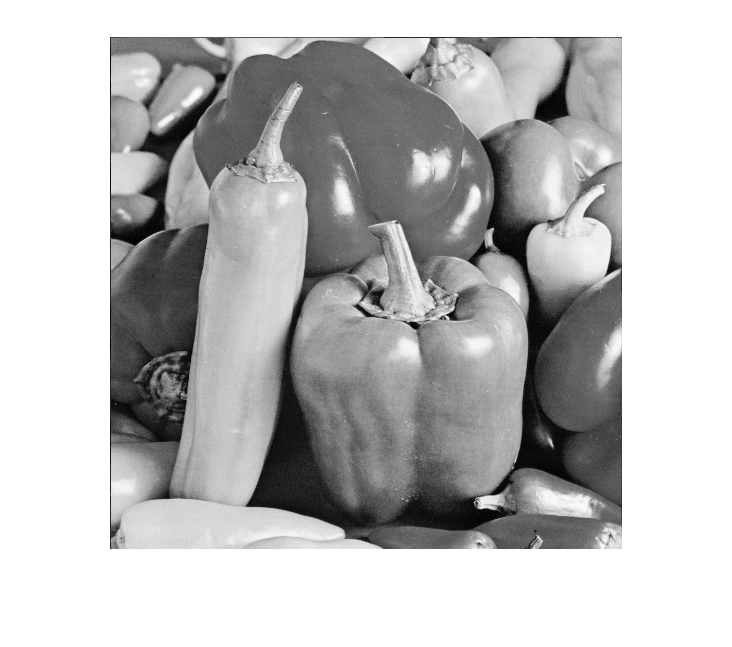}} \\
			\subfloat[Without filtering, PSNR=$18.80$~dB]{\includegraphics[trim=2.6cm 2cm 2.6cm 2cm, clip=true, width=1.8in]{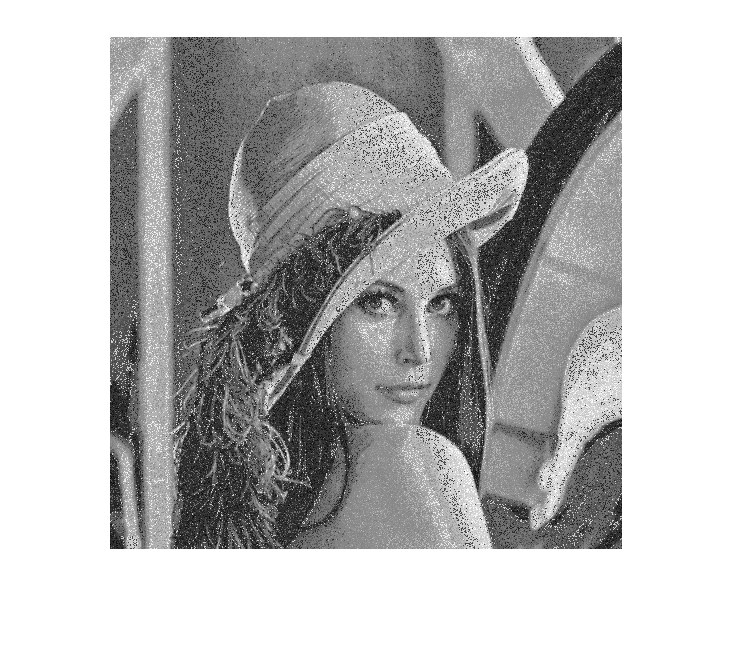}} & \subfloat[Without filtering, PSNR=$18.42$~dB]{\includegraphics[trim=2.6cm 2cm 2.6cm 2cm, clip=true, width=1.8in]{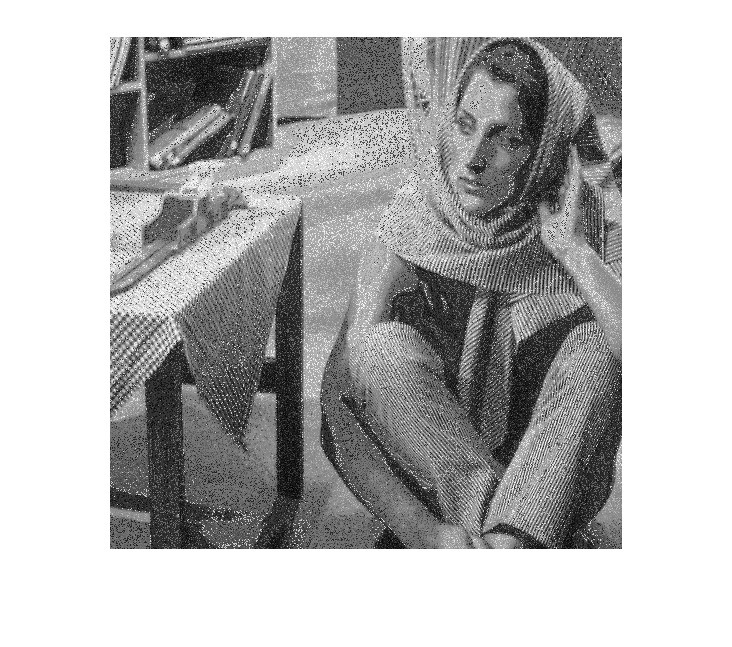}} & \subfloat[Without filtering, PSNR=$18.85$~dB]{\includegraphics[trim=2.6cm 2cm 2.6cm 2cm, clip=true, width=1.8in]{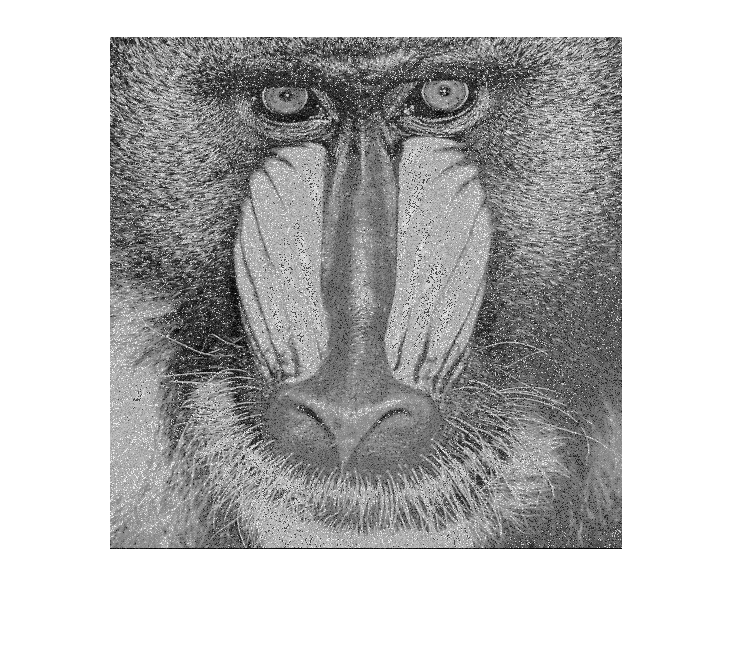}} &\subfloat[Without filtering, PSNR=$18.64$~dB]{\includegraphics[trim=2.6cm 2cm 2.6cm 2cm, clip=true, width=1.8in]{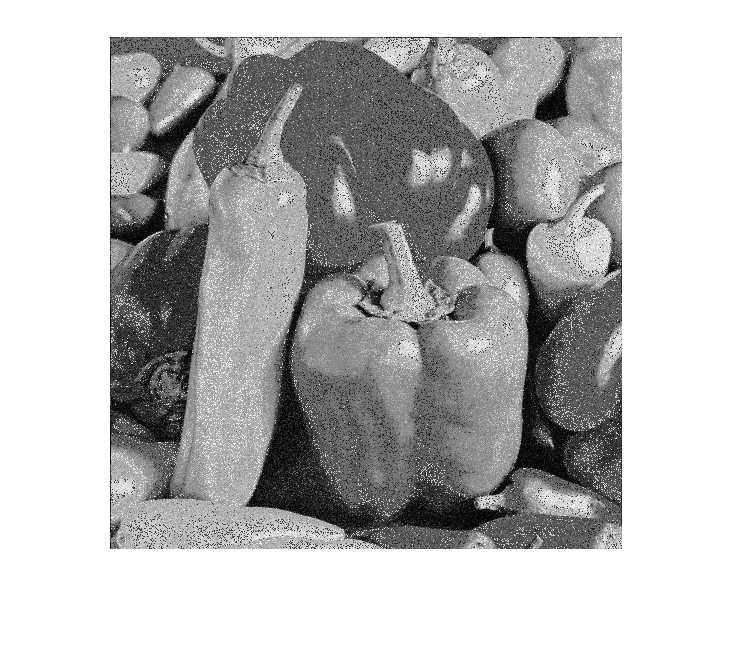}}
			\\
			\subfloat[Gaussian filter, PSNR=$24.40$~dB]{\includegraphics[trim=2.6cm 2cm 2.6cm 2cm, clip=true, width=1.8in]{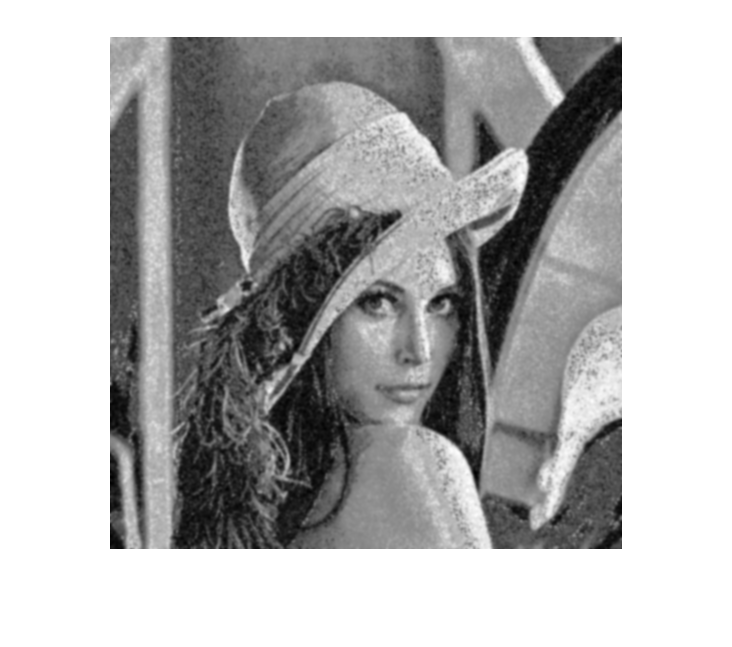}} & \subfloat[Gaussian filter, PSNR=$22.51$~dB]{\includegraphics[trim=2.6cm 2cm 2.6cm 2cm, clip=true, width=1.8in]{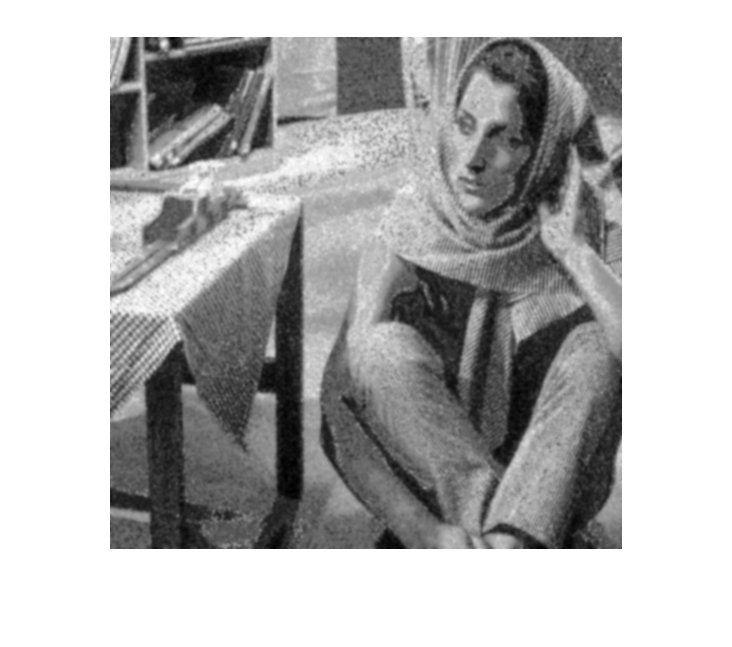}} & \subfloat[Gaussian filter, PSNR=$21.51$~dB]{\includegraphics[trim=2.6cm 2cm 2.6cm 2cm, clip=true, width=1.8in]{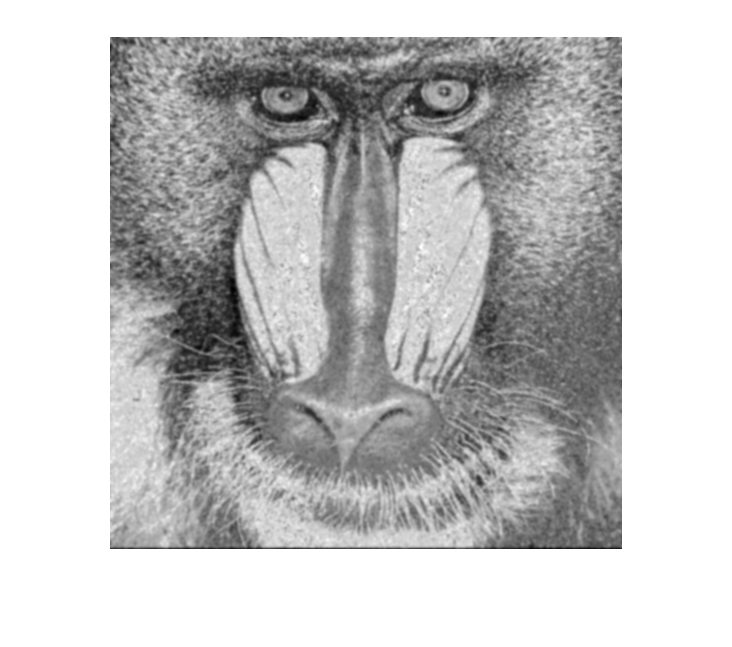}} & \subfloat[Gaussian filter, PSNR=$24.89$~dB]{\includegraphics[trim=2.6cm 2cm 2.6cm 2cm, clip=true, width=1.8in]{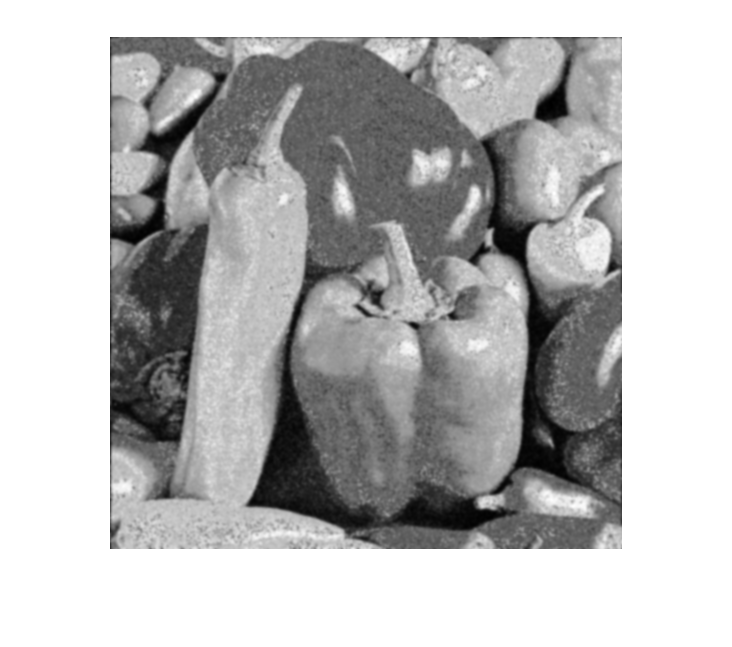}}\\
			\subfloat[BM3D filter, PSNR=$26.48$~dB]{\includegraphics[trim=2.6cm 2cm 2.6cm 2cm, clip=true, width=1.8in]{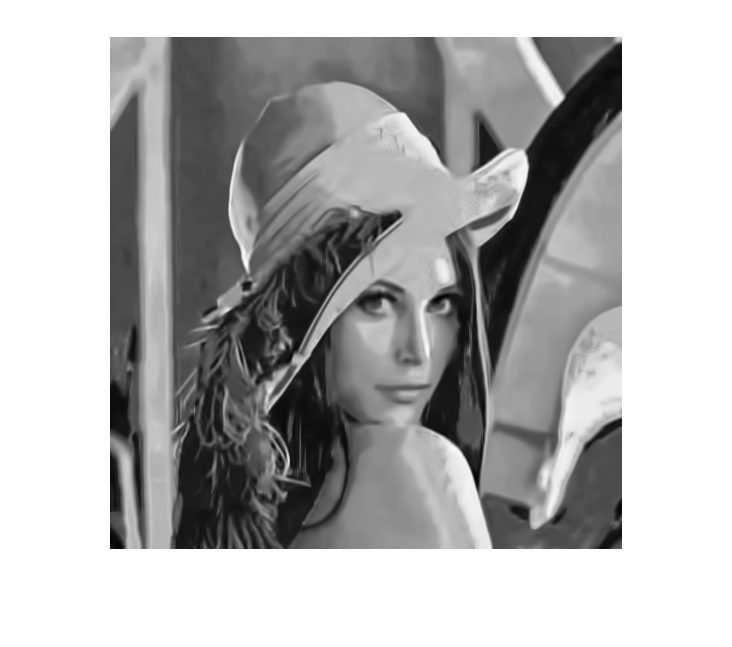}} & \subfloat[BM3D filter, PSNR=$25.64$~dB]{\includegraphics[trim=2.6cm 2cm 2.6cm 2cm, clip=true, width=1.8in]{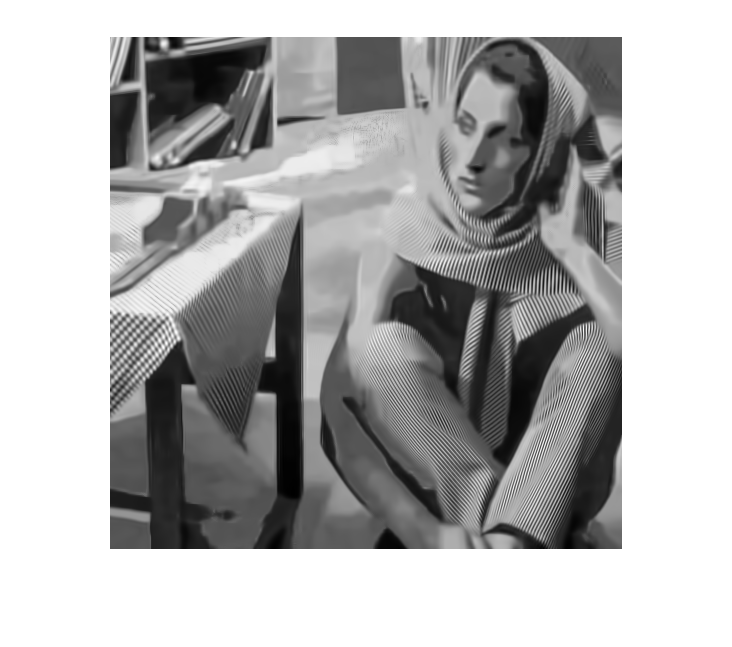}} & \subfloat[BM3D filter, PSNR=$22.22$~dB]{\includegraphics[trim=2.6cm 2cm 2.6cm 2cm, clip=true, width=1.8in]{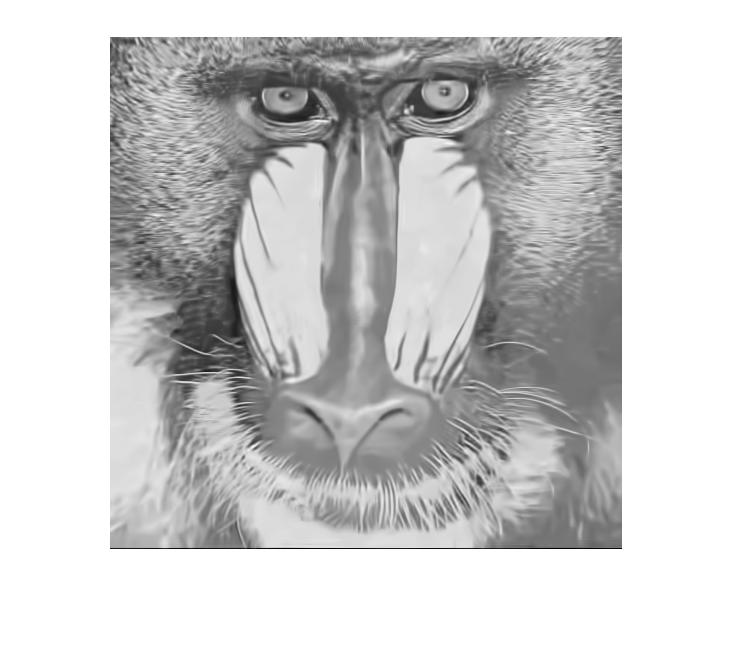}} & \subfloat[BM3D filter, PSNR=$26.17$~dB]{\includegraphics[trim=2.6cm 2cm 2.6cm 2cm, clip=true, width=1.8in]{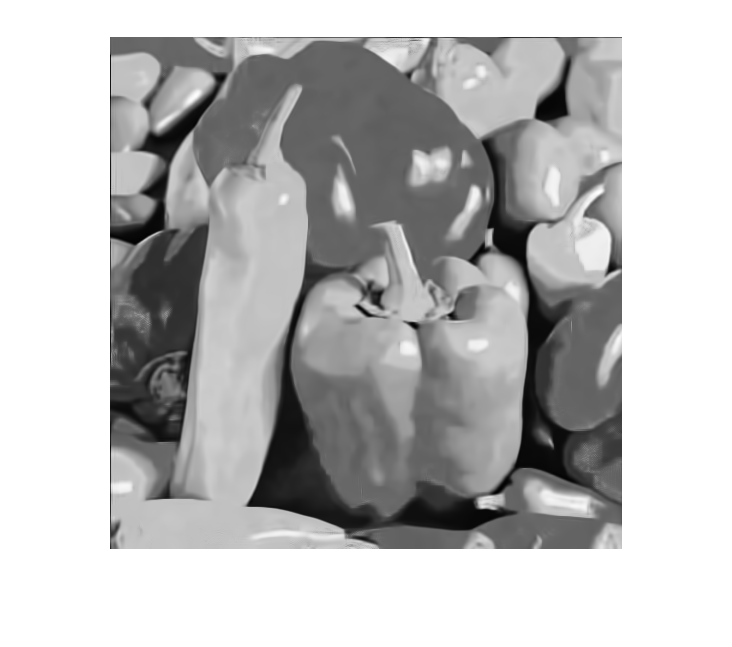}} 
	\end{tabular}}
	\caption{The visual quality of recovered images from different recovery methods: The first row contains four original images. The second row contains recovered images by Algorithm~\ref{Algorithm1} but without post-filtering. The third row contains recovered images by Algorithm~\ref{Algorithm1} with a Gaussian filter. The last row contains recovered images by Algorithm~\ref{Algorithm1} with a BM3D filter.} \label{fig1}
\end{figure*}
\begin{figure}[t]
	\centering
	\includegraphics[trim=0.2cm 0.0cm 1.1cm 0.6cm, clip=true, width=3.3in]{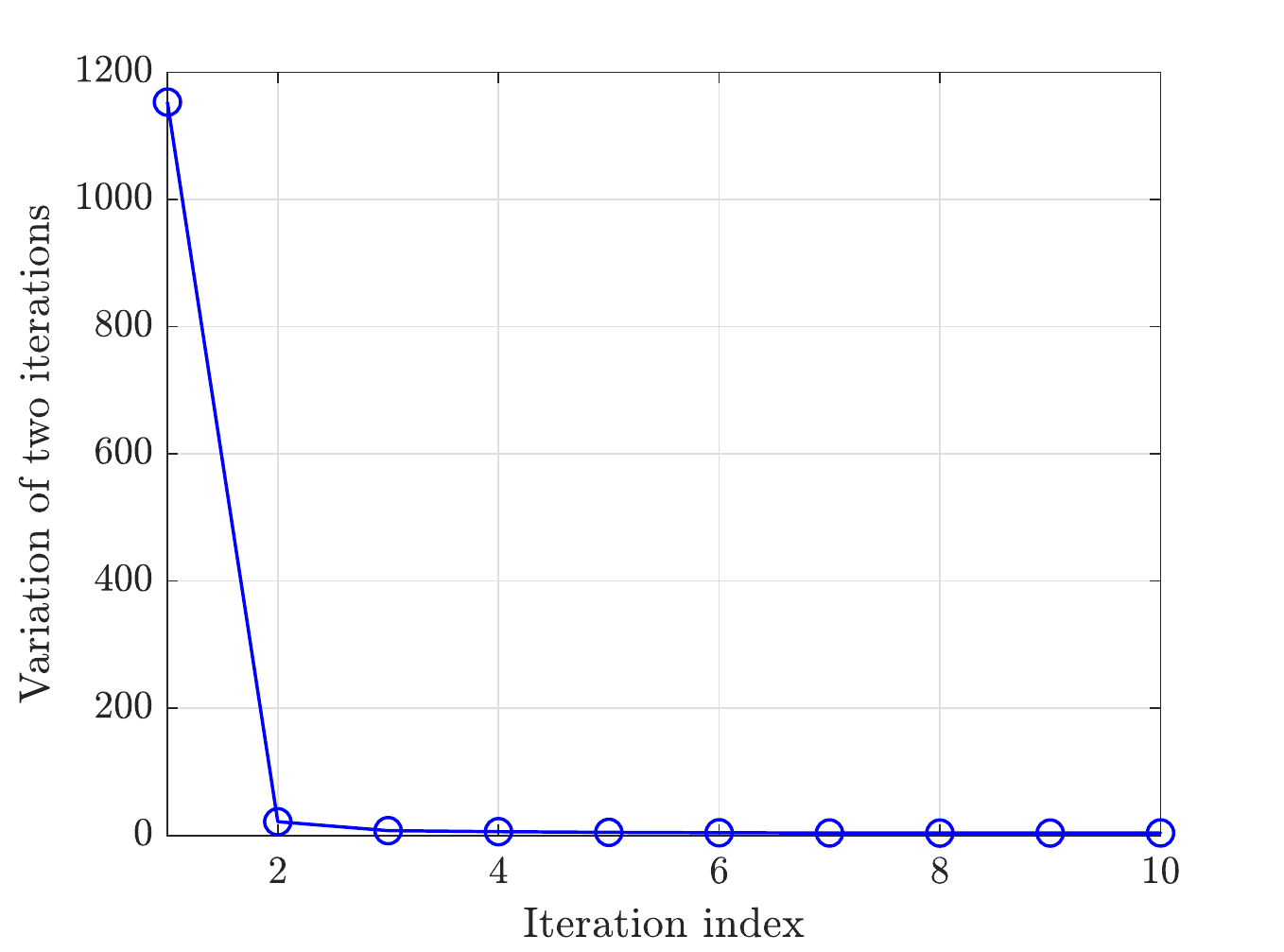} 
	\caption{The convergence of ADMM approach.}
	\label{FigConvergenence}
\end{figure}
where $\nu_z > 0$ is a regularization parameter. We notice that the solution to $\pmb{\alpha}_z$ is not unique and it is heavily based on each dictionary $\mathcal{D}$. The solution to problem~\eqref{Probv2} and an effective way to define $\mathcal{D}$ based on each noisy image $\widehat{\mathbf{X}}$ is presented similarly as in, for example the BM3D filter \cite{dabov2007image} or the K-SVD filter \cite{aharon2006k}. The use of a post filter to mitigate the remaining noise and artifacts in recovered images is integrated with the previous ADMM algorithm as shown in Algorithm~\ref{Algorithm1}.
\section{Numerical Results}
We consider a Massive MIMO network where the base station and user are equipped with $64$ and $4$ antennas, respectively. The $256-$QAM is used to map image data into the constellation points. The SNR level is $5$~dB. The parameter setup of the post filters are selected to get the best filtering results in term of the peak signal-to-noise ratio (PSNR) [dB]. In particular, for Gaussian filter, the standard derivation is $1.3$ and that figure of the BM3D filter is $41$. The small-scale fading follows the normal symmetric complex Gaussian distribution. The initial version of $\mathbf{s}_z$ is set as $\mathbf{s}_z^{(0)} = \mathbf{H}^H \mathbf{y}$. We select $ \beta_z = 1.26$ and $\alpha = 1.62$. The initial Lagrange multiplier value is zero, i.e., $\pmb{\mu}_z^{(0)} = \mathbf{0}$. Algorithm~\ref{Algorithm1} is testified by utilizing four different natural images comprising Lenna, Barbara, Mandrill, and Peppers. 

Fig.~\ref{fig1} visualizes the recovered quality of the proposed algorithm by utilizing different images. It shows that reconstructing image data information without taking the correlation among pixels results in the remaining of noise and artifacts as shown in the second row of Fig~\ref{fig1}, especially for the Lenna and Barbara images. Consequently, the PSNR values of those recovered images are only 18.80 dB and 18.42 dB, respectively. Meanwhile, Mandrill and Peppers recovered quality are slightly better than two previous images with the PSNR values of 18.85 dB and 18.64 dB, respectively. By exploiting a Gaussian filter to subtract remaining noise and artifacts, we observe significant improvements of the refined images. For the Lenna image, the PSNR value gets $29.8\%$ better when comparing with not using a post filter. The corresponding improvement of the Peppers image is up to  $33.53\%$. The last row of Fig.~\ref{fig1} demonstrates the effectiveness of an advanced post filter by exploiting the non-local image information. The refined images have better quality in both PSNR and visualization, which are closer to the ground truth. These results manifest the importance of utilizing a proper post filter in enhancing restored images quality.

Fig.~\ref{FigConvergenence} shows the convergence in Algorithm~\ref{Algorithm1} used the cost function in \eqref{eq:Stopping}. The result is averaged over different image data column vectors. It demonstrates that the ADMM approach converges very fast, just after $3$ iterations. As aforementioned, Algorithm~\ref{Algorithm1} has low computational complexity per each iteration due to the low cost of updating each sub-problem, therefore, this algorithm can serve the purposes of image restoration.

\section{Conclusion}
This paper has suggested an optimization solution and post-filtering framework to recover transmitted image data in Massive MIMO communications. The optimization solution based on the alternating direction method of multipliers approach, utilized in order to have an initial version of the decoded image at the receiver. We then demonstrated that the exploitation of a low pass filter can refine the quality of images effectively. Our framework is immensely important for multi-media applications in 5G-and-beyond systems. The future works should include an image structure constraint in the optimization problem for better reconstruction and the use of artificial intelligence for real-time processing purposes.

\bibliographystyle{IEEEtran}
\bibliography{IEEEabrv,refs}
\end{document}